# Ferroelectric-ferroelastic phase transition in a nematic liquid crystal

Nerea Sebastián[1], Luka Cmok[1], Richard J. Mandle[2], María Rosario de la Fuente[3], Irena Drevenšek Olenik[1,4], Martin Čopič[1,4], and Alenka Mertelj[1]

[1]*J. Stefan Institute, SI-1000 Ljubljana, Slovenia*
[2]*Department of Chemistry, University of York, York, YO10 5DD, UK*
[3]*Department of Applied Physics II, University of the Basque Country (UPV/EHU), Apdo.644-48080 Bilbao, Spain*
[4]*Faculty of Mathematics and Physics, University of Ljubljana, SI-1000 Ljubljana, Slovenia*

(Date: 2019-10-09)

Ferroelectric ordering in liquids is a fundamental question of physics. Here, we show that ferroelectric ordering of the molecules causes formation of recently reported splay nematic liquid-crystalline phase. As shown by dielectric spectroscopy, the transition between the uniaxial and the splay nematic phase has the characteristics of a ferroelectric phase transition, which drives an orientational ferroelastic transition via flexoelectric coupling. The polarity of the splay phase was proven by second harmonic generation (SHG) imaging, which additionally allowed for determination of the splay modulation period to be of the order of 5 - 10 microns, also confirmed by polarized optical microscopy. The observations can be quantitatively described by a Landau-de Gennes type of macroscopic theory.

**Subject Areas:** Soft Matter, Materials Science, Phase transitions

A ferroelectric fluid, and more specifically a ferroelectric nematic phase, is of great fundamental and practical interest. Nematic phases retain truly 3D fluidity as their molecules exhibit only orientational but no positional order. Molecular shape is of key importance on determining the occurrence of the different mesophases. While highly symmetrical rod-like molecules give rise to a uniaxial nematic phase, complex molecular shapes may promote nematic phases with higher orientational order as for example biaxial [1] or twist-bend nematic phases [2] . On the other hand, chiral molecules, beside the chiral nematic phase, also form three "blue" phases. Although theoretical works anticipate that polar order is more likely to occur in systems made of disk-like constituents [3,4], for those consisting of pear- or wedge-shaped molecules, softening of the splay elastic constant was predicted [5] and the occurrence of splayed polar nematic phase was theoretically demonstrated [6,7]. Although the possibility of a ferroelectric nematic phase was already envisioned by Born [8], experimental evidence has only been observed in some polymer liquid crystals [9], while in low molecular mass materials it has until now eluded experimental discovery.

The so called splay nematic phase was just recently discovered [10,11] to appear in materials made of molecules with a large dipole moment and a lateral group, which results in a slightly wedge molecular shape (Fig.1a) [12,13]. These materials exhibit two distinct nematic mesophases, the usual uniaxial nematic phase and the lower temperature splay nematic phase. The weakly first order phase transition between the phases is accompanied by a significant pretransitional behavior, manifested as strong splay orientational fluctuations. The transition is driven by instability towards splay orientational deformation, which leads to a periodically splayed structure. Because the splay nematic phase in planar confinement appears optically homogeneous, it was speculated that the modulation period of the splay phase was of nm size similarly as the modulation of the twist bend nematic phase [14,15].

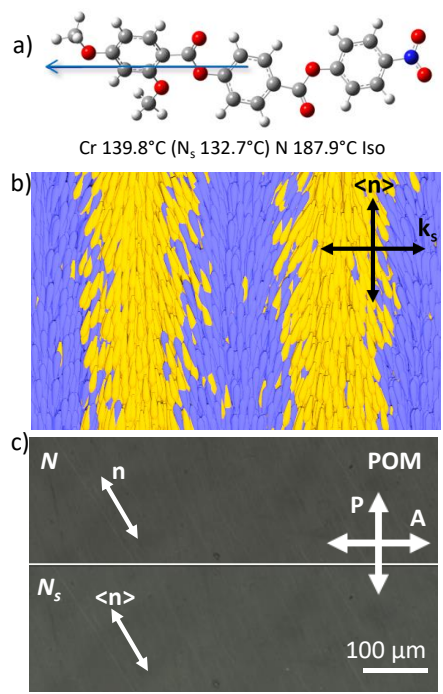

FIG. 1. (Color online) (a) Schematic of a molecule of RM734 and its phase sequence. (b) Schematic of the splay phase, where different colors denote molecular orientation either along **n** or –**n**. $\mathbf{k}_s$ denotes the splay wave vector. (c) POM images of RM734 in a homogeneous in-plane alignment cell in $N$ and $N_s$ phases (thickness 20 μm) cell.

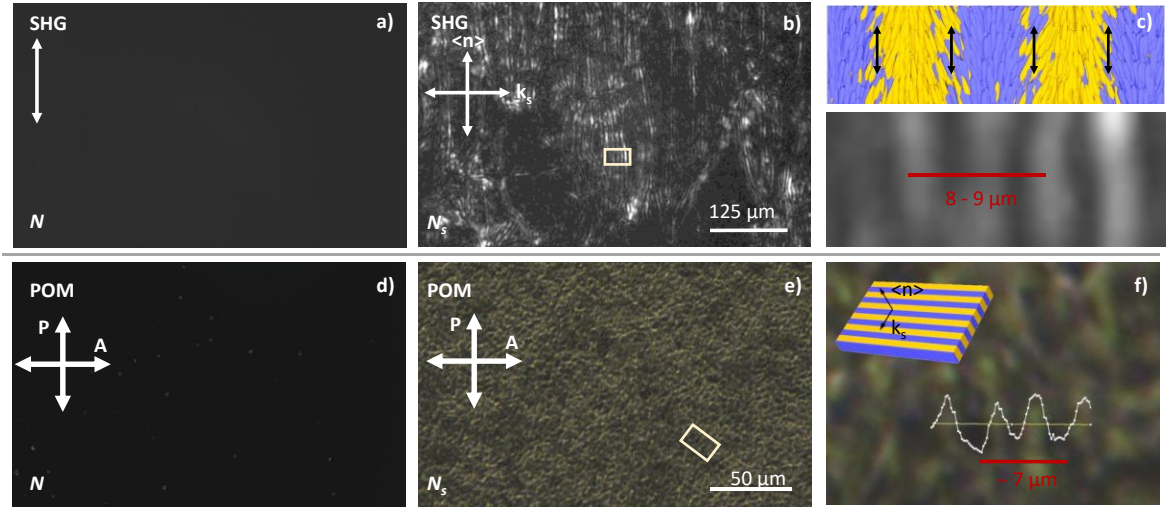

FIG. 2. (Color online) (a-c) SHG microscopy images in the a) $N$ phase ($T_{NNs}$ ~ 2 K) and b) transition to $N_s$ phase ($T_{NNs}$ ~ -0.5 K) in a planar cell (thickness 20 μm). An enlarged section from b) is shown in c) and compared with the proposed structure. (d-f) Polarization microscopy images in the d) $N$ ($T_{NNs}$ ~ 10 K) phase and e) $N_s$ phase ($T_{NNs}$ ~ -10K) for a cell with bare ITO electrodes (thickness 20 μm). An enlarged section from e) is shown in f) where the white dots indicate the mean intensity across that section. Rectangles b) and e) correspond to the enlarge images in c) and f) respectively.

Here, we experimentally demonstrate that the splay modulation is of the order of microns and that the phase transition to the splay nematic phase is a ferroelectric-ferroelastic transition, in which flexoelectric coupling causes simultaneous occurrence of diverging behavior of electric susceptibility and of instability towards splay deformation.

The usual uniaxial nematic phase ($N$) is non-polar, and the average orientation of the molecules is described by a unit vector - director **n -,** which has inversion symmetry $\mathbf{n} \equiv -\mathbf{n}$. In the splay nematic phase ($N_s$), the director exhibits a modulated splayed structure (Fig.1b), which is biaxial and can be described by the average orientation of the director $\langle \mathbf{n} \rangle$, the splay wave vector $\mathbf{k}_s$, and the modulation period $\Lambda = 2\pi / |\mathbf{k}_s|$. Additionally, the splay nematic phase exhibits electric polarization **P**, whose magnitude oscillates along the splay direction. In LC cells with homogeneous in-plane alignment, well oriented sample optically looks very similar in both phases (Fig. 1c) [1].

As SHG can only be observed in structures with no center of inversion symmetry, in particular in polar structures, it is a very sensitive technique to detect system polarity. Thus, we proved the appearance of polar order during the $N$-$N_s$ transition by SHG microscopy. In the experiment, a layer of RM734 in a LC cell (Instec Inc., thickness 20 μm), is illuminated by pulsed laser light (λ = 800 nm, pulse length: 100 fs, repetition rate: 1 kHz) and imaged with a system composed of a 20X microscopy objective and a CMOS camera (detailed description in the SI).

In the $N$ phase, SHG image appears completely dark, as expected for a homogeneously oriented non-polar nematic liquid crystal (Fig 2(a)). On cooling, the transition to the splay phase is characterized by a sudden strong SHG signal (coinciding with the destabilization of homogeneous orientation observed by polarization microscopy (POM) [10]). During the transition, the SHG image clearly shows a periodic stripped texture (Fig 2(b)) with a periodicity of around 8 to 9 µm perpendicular to the rubbing direction. In agreement with the proposed phase structure, the bright regions correspond to the splayed deformation and the dark lines to those where **P** = 0 (Fig 2(c)). The SHG activity clearly demonstrates that the $N_s$ phase is polar. Right below the $N-N_s$ transition, the splay wave vector orients perpendicularly to the plane of the sample and the periodic structure is only observed around defects (SI Fig. S1(c)).

Interestingly, while in the cells treated for perpendicular alignment of **n,** random in-plane alignment was obtained [10], in non-treated ITO cells, **n** in the $N$ phase aligns out of plane (Fig 2.d). In the $N_s$ phase, the $\mathbf{k}_s$ lays then in the plane of the sample, but there is no preferred in-plane orientation for it, so its direction changes with position, resulting in a disordered structure discernable after the phase stabilization. In a POM image, small regions having a regular modulation with a periodicity of about 7 µm are observed (Fig 2.(e-f)), in correspondence with SHG observations in the transition and around defect lines.

To gain insight into the polarization mechanisms involved in the $N$-$N_s$ phase transition, we measured the complex permittivity $\varepsilon(\omega) = \varepsilon'(\omega) - i\varepsilon''(\omega)$ in the frequency range 10 Hz – 110 MHz (See SI for detailed description). The sample consisted of two circular gold-plated brass electrodes of 5 mm diameter and separated by 50 µm thick silica spacers. The untreated gold surfaces spontaneously gave rise to perpendicular alignment, and thus, parallel component of the permittivity was measured (Fig 3.a). The measuring oscillator level was set to 10 mV$_{rms}$ and measurements were performed on cooling. The probe electric field was low enough for the measurements to be in the linear regime in the $N$ phase. However, in the $N_s$ phase, optical switching was detected for fields as low as the employed probe field, and thus, interpretation of results is challenging and will be reported elsewhere.

Characteristic frequency and amplitude of each process are obtained by fitting $\varepsilon(\omega)$ to the Havriliak-Negami (HN) equation [16] (SI for fit examples) (Fig 3). In the isotropic phase ($I$), a single relaxation process ($m_{iso}$) is detected with frequency of the order of 20 MHz for the rigid dipolar molecule rotating in the isotropic environment. $\Delta\varepsilon_{iso}$~17 is in agreement with the large calculated molecular longitudinal dipole moment ~11.4 D [10]. In the $N$ phase, immediately after the $I$-$N$ transition and in the frequency range studied, a single process $m_{\parallel,1}$ is detected with lower frequency and larger amplitude than in the $I$ phase. On cooling $\Delta\varepsilon_{\parallel,1}$ steadily increases and a broadening of the relaxation can be detected. Far below $T_{IN}$ it is possible to resolve such broadening into a second process $m_{\parallel,2}$ with lower strength ($\Delta\varepsilon_{\parallel,2}$ ~40), which becomes clear by analyzing $d\varepsilon'(f)/d(\log(f))$ vs $\log(f)$ (SI). It appears that close to $T_{IN}$ the detected mode, in line with usual nematic behavior for molecules with longitudinal dipole moment, can be attributed to molecular rotations around the short molecular axis. However, collective behavior starts to develop on lowering the temperature, as evidenced by the strength increase. As collective motions slow down faster than the molecular process, at around 160 °C, both processes become distinguishable, being then possible to attribute $m_{\parallel,2}$ to the molecular rotations around the short molecular axis and $m_{\parallel,1}$ to a collective movement of dipole moments. Additionally, a third mode $m_{\parallel,3}$ is detected in the higher part of the frequency range (~ 1-10 MHz). The strength of this mode is very small compared to the main contributions, and thus both, its frequency and its amplitude as depicted in Fig. 3(b) and Fig. SI6, are just tentative. This mode, by frequency and amplitude, can be attributed to the rotation around the molecular long axis, as described by the Nordio-Rigatti-Segre theory [17]. On further approaching the $N$-$N_s$ phase transition, softening of the $m_{\parallel,1}$-mode is evidenced by the strong decrease of the relaxation frequency and divergence of $\Delta\varepsilon_{\parallel,1}$, as the result of molecular motions becoming collective. Complete condensation of the mode could not be observed as, close to the transition, $m_{\parallel,1}$ is masked by conductivity effects and can only be followed until two degrees before the transition, for which $f \sim 800$ Hz.

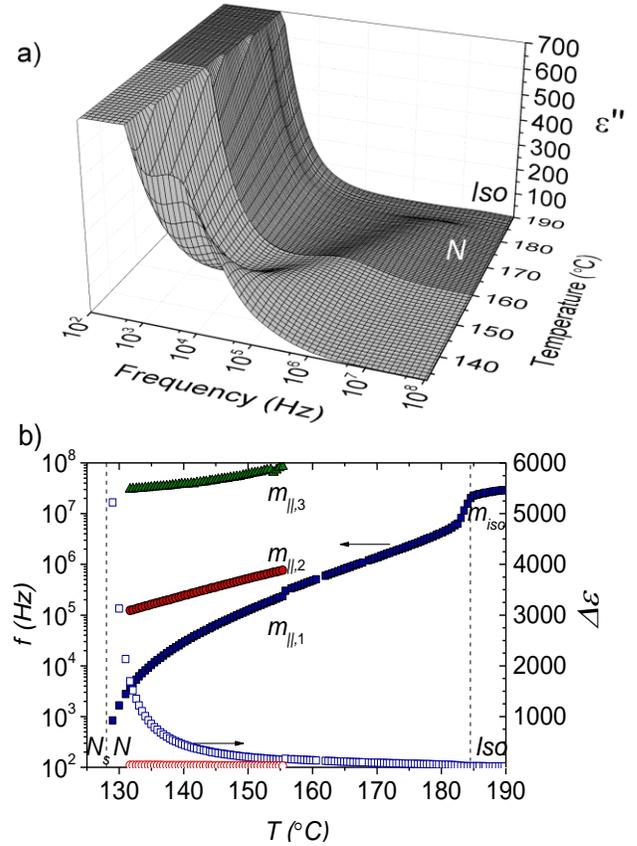

FIG. 3. (Color online) a) 3D plot of the dielectric losses $\varepsilon''$ vs temperature and frequency. b) Temperature dependence of the relaxation frequencies and strengths of the different processes in the isotropic and nematic phases.

Theoretically, this transition between $N$ and $N_s$ phases can be described by a Landau-de Gennes type of the free energy [10,18]. The order parameter of the nematic phase is a tensor and to understand the temperature behavior of the elastic constants on a larger temperature interval a tensor description is necessary [18]. However, in the case of the $N$ - $N_s$ transition, the critical behavior is limited to the temperature range where the scalar order parameter is approximately constant [11], and the description of the nematic phase by $\mathbf{n}$ is sufficient. In the $N_s$ phase, as demonstrated experimentally by SHG, the inversion symmetry is broken, which means that an additional vector order parameter is needed. If molecules have dipole moments with a component along the symmetry axis, as is in our case, then the vector order parameter is the electric polarization $\mathbf{P}$. However, in general, the origin of the order parameter vector is not necessarily the electric polarization, but can be for example the shape of the molecule [5,19]. In the $N$ phase, splay or bend orientational deformations give rise to flexoelectric polarization, which is the result of the coupling between $\mathbf{P}$ and the deformation of $\mathbf{n}$ [20]. It has been shown that this coupling can be a driving mechanism for a transition from the uniaxial to a modulated nematic phase [10,19,21]. The free energy density with minimum number of terms necessary for the description of the $N$ - $N_s$ transition is then

$$f = \tfrac{1}{2}K_1(\nabla\cdot\mathbf{n})^2 + \tfrac{1}{2}K_3(\mathbf{n}\times(\nabla\times\mathbf{n}))^2 - \gamma\mathbf{n}(\nabla\cdot\mathbf{n})\cdot\mathbf{P} + \tfrac{1}{2}t\mathbf{P}\cdot\mathbf{P} + \tfrac{1}{2}b(\nabla\mathbf{P})^2 \quad (1)$$

Here, the first two terms are the usual Frank elastic free energy terms, where $K_1$ and $K_3$ are the splay and the bend orientational elastic constants, respectively. The third term describes the coupling between the splay deformation and the electric polarization, where $\gamma$ is a bare splay flexoelectric coefficient. The last two terms are the lowest by symmetry allowed terms in $\mathbf{P}$ and $\partial P_i/\partial x_j$. In our case, the molecular dipole moment is along the molecular long axis, so $\mathbf{P} = P\mathbf{n}$. Although the $N$-$N_s$ phase transition described by Eq. (1) is a second order phase transition, it can be extended to describe a weakly first order phase transition by adding higher order terms [22].

Because of the flexoelectric coupling, the orientational fluctuations of $\mathbf{n}$ are coupled with the fluctuations of $\mathbf{P}$, which causes the existence of two branches of the splay eigenmodes (see SI for details): a hydrodynamic mode, which is mostly a director mode, and an optic mode, which is predominantly a polarization mode. By dielectric spectroscopy, the polarization eigenmode at $q = 0$ is measured. Its relaxation rate is $t/(2\eta_P)$, where $\eta_P$ is the dissipation coefficient for $\mathbf{P}$, and the square amplitude of fluctuations is $\langle P^2\rangle = 2k_BT/(Vt)$, where $V$ is the volume of the sample. $\langle P^2\rangle$ is proportional to $\Delta\varepsilon_{\parallel,1}$ (see SI) [23].

The flexoelectric coupling in general causes that the splay elastic constant measured in the experiments is rescaled as $K_{1eff} = K_1 - \gamma^2/t$ (see SI and [10,18]). If polar order is not favorable, then $t$ is large and the flexoelectric coupling barely affects the measured value of $K_1$. However, when approaching the ferroelectric phase transition, $t$ decreases with temperature and, in a usual Landau description, the transition to a ferroelectric phase would happen when $t$ becomes negative [22]. However, if the flexoelectric coupling exists, then the phase will become unstable towards splay deformation at a positive, critical value $t_c = \gamma^2/K_1$, i.e., before $t$ becomes negative. In general, the coefficient $t$ is related to the electric susceptibility $\chi$, as $t = 1/(\varepsilon_0\chi)$, where $\varepsilon_0$ is the vacuum permittivity. In our case, the collective polarization $m_{\parallel,1}$-mode is the one that drives the ferroelectric transition and its contribution to the susceptibility determines $t = \left(\varepsilon_0 \Delta\varepsilon_{\parallel,1}\right)^{-1}$ (see SI), so the effective splay elastic constant can be written

$$K_{1,eff} = K_1 - \varepsilon_0 \gamma^2 \Delta\varepsilon_{\parallel,1} \quad (2)$$

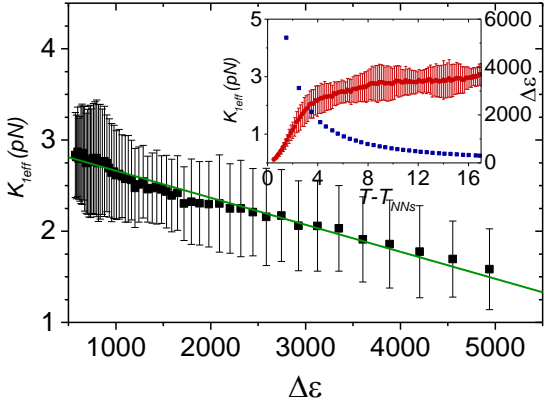

FIG. 4. (Color online) Measured values of $K_{1,eff}$ vs interpolated values of $\Delta\varepsilon_{\parallel,1}$. Inset shows the temperature dependence of both, $K_1$ and $\Delta\varepsilon_{\parallel,1}$, few degrees above the transition to the $N_s$ phase. The green line corresponds to the linear fit from which $\gamma = 0.006$ V is obtained.

The dependence of the measured effective splay elastic constant [10] on $\Delta\varepsilon_{\parallel,1}$ (Fig. 4) substantiates the linear dependency predicted by Eq. (2). The linear fits yields the values of $\gamma = 0.006$ V , $K_1 = 3$ pN and, $\varepsilon_0 t_c = 10^{-4}$ .

The modulation period just below the $N$-$N_s$ phase transition is determined by the ratio of the elastic constants $K_1$ and $b$, and the flexoelectric coefficient: $\Lambda = 2\pi\sqrt{3bK_1/\left(\gamma^2 \Delta t_n\right)}$ , where $\Delta t_n = 1 - t/t_c$ [10].

If flexoelectric coupling is large, then the modulation period is small and vice versa. In our case, the value $\gamma$ is small and so the modulation period is in the µm range. In this size range, it should be visible in POM imaging, however, there are several effects that hinder such observation. The SHG imaging has shown that in LC cells treated for homogeneous in-plane alignment, after the stabilization of the phase, the splay wave vector tends to align perpendicularly to the plane of the cell, and a periodic structure is only seen around defects. In polarizing microscopy, such periodic structure around defects cannot be distinguished (SI Fig. S1.c). A possible reason is that the splay deformation is so small that the scattering from orientational fluctuations, which are in such geometry strongly expressed, masks it. Additionally, the deformation of **n** close to the defect causes diffraction of light, which again hinders POM imaging. In cells without alignment layer, in which for the $N$ phase, **n** aligns perpendicularly to the sample plane, in the $N_s$ phase, the modulation is visible (Fig. 2 (e)-(f)) as this case corresponds to the geometry in which fluctuations do not cause scattering of light [24]. Observed periodicity and sample confinement are of the same order and consequently, determination of the temperature dependence of the periodicity requires additional methods. However, our observations suggest that it remains of the same order down to few degrees below the transition. In agreement with these observations, splay modulation period in the order of µm was also very recently observed optically in a two-component mixture [25]. It should also be mentioned that a phase showing similar textures and polarity was recently observed in a liquid crystal compound with a 1,3-dioxane unit [26].

Here it is important to note that the ferroelectric phase transition described by Eq. (1) will always be accompanied by a ferroelastic transition even if the flexoelectric coupling is very small. This case is very similar to ferroelectric–ferroelastic phase transitions in some solid materials, in which linear coupling between strain and polarization is by symmetry allowed, as for example in KDP [27]. The main difference between the transition in a solid or in a nematic liquid crystal is in the deformation type, which is a strain in the former case, while in the latter, it is orientational. While strain can be homogeneous, it is not possible to fill the space with homogeneous splay, and consequently, a modulated structure is formed.

In conclusion, we have experimentally shown that in the recently discovered splay nematic phase, the modulation period is of macroscopic size. The transition from the uniaxial to the splay nematic phase is a ferroelectric phase transition, which is evidenced by the divergent behavior of electric susceptibility, and is, due to the flexoelectric coupling, accompanied by orientational ferroelastic transition. The observed behavior rises fundamental questions about polar order in liquids, as for example, what combination of molecular charge distribution and shape is necessary that the molecules will on average orient in the same direction while still remain in a liquid state. And, would any of such combinations allow for realization of a homogeneous uniaxial ferroelectric nematic phase or is it essential that ferroelectric ordering in a nematic liquid crystal is accompanied by orientational deformation.

NS, IDO, MČ and AM acknowledge the financial support from the Slovenian Research Agency (research core funding No. P1-0192). LC acknowledge support from MIZŠ&ERDF funds OPTIGRAD project, 2014-2020. RJM acknowledges QinetiQ for the award of an ICASE studentship.

See Supplemental Material at [*URL will be inserted by publisher*].

# Supplementary information to
# Ferroelectric-ferroelastic phase transition in a nematic liquid crystal

Nerea Sebastián[1], Luka Cmok[1], Richard J. Mandle[2], María Rosario de la Fuente [3], Irena Drevenšek Olenik[1,4], Martin Čopič[1,4], and Alenka Mertelj[1]

[1]*J. Stefan Institute, SI-1000 Ljubljana, Slovenia*
[2]*Department of Chemistry, University of York, York, YO10 5DD, UK*
[3]*Department of Applied Physics II, University of the Basque Country (UPV/EHU), Apdo. 644, 48080 Bilbao, Spain*
[4]*Faculty of Mathematics and Physics, University of Ljubljana, SI-1000 Ljubljana, Slovenia*

(Date: 2019-10-09)

### A. Material

RM734 was synthesized according to Ref. [1] and additionally purified as described in Ref. [2]. As determined by means of differential scanning calorimetry phase transition temperatures have been determined to be: isotropic to nematic $T_{IN}$= 187.9 °C, nematic to second nematic transition at $T_{NNs}$ = 132.7 °C, and a melting point at $T_m$ = 139.8 °C. We used the Gaussian G09e01 suite of programs [3] to determine the B3LYP/6-31G(d) minimized geometry of RM734 shown in Fig.1. The molecular dipole moment calculated by the B3LYP/6- 31G(d) level of DFT is of 11.3748D and oriented almost along the molecular long axis.

### B. Experimental Methods

**B.1 Liquid Crystal cells**

For optical observations commercially available glass cells have been used, which consists of two transparent glass plates between which the liquid crystal is introduced by capillarity. Both glass plates are lithographed with a transparent conductive layer of Indium Tin Oxide (ITO). To achieve well defined directions of the molecular director at the cell interface, the electrode surfaces are usually treated with chemical agents. For thin enough cells, this order would be transmitted throughout the cell. Two different alignments can be distinguished depending on the angle formed by the molecular director and the surface. If the angle is zero, we have a planar alignment, in which the molecular director is parallel to the surface (in-plane alignment). In contrast, we have an homeotropic alignment when the molecular director is perpendicular to the cell surface (out of plane alignment). Due to the high resistance of the ITO layer, the cells behave as an RC series circuit and thus, a spurious absorption appears in the dielectric spectra, whose characteristic frequency will depend on the dielectric properties of the studied material. This fact limits the use of ITO glass cells in frequency and thus, as will be described below, specialized cells have been used for the dielectric measurements.

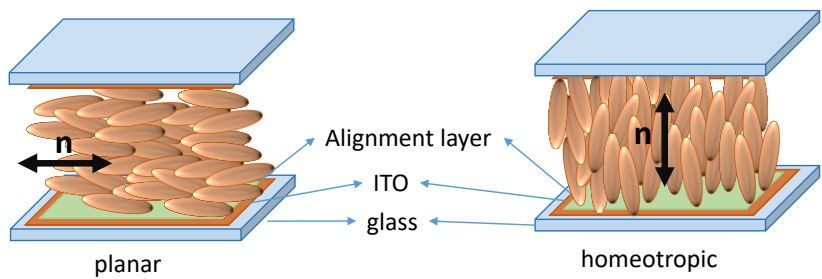

Fig. SI.1 Schematic representation of liquid crystal cells with planar and homeotropic alignment.

**B.2 Second Harmonic Imaging**

SHG imaging was preformed using an ultrafast Ti:Sapphire amplifier laser source (Legend Elite, Coherent) operating at λ = 800 nm with 100 fs pulse length and 1 kHz repetition rate. Probing beam intensity and polarisation was tuned with a system of polarizers followed by a quarter-wave plate. Incident beam on the sample was circularly polarized and the power was selected, below the damage threshold of the sample. Sample in a LC cell (20 μm planar Instec Inc.) was placed in a heating stage (Instec HCS412W) and controlled by temperature controller (mK2000, Instec). The spot size of the laser beam on the sample was around 2 mm. Image acquisition was done by custom build microscope system (Nikon SLWD 20x objective) with resolving power of 0,3 μm/pixel. Fundamental radiation was blocked internally by two high pass filters before image was recorded using a CMOS camera (FLIR Blackfly BFLY-U3-23S6M-M).

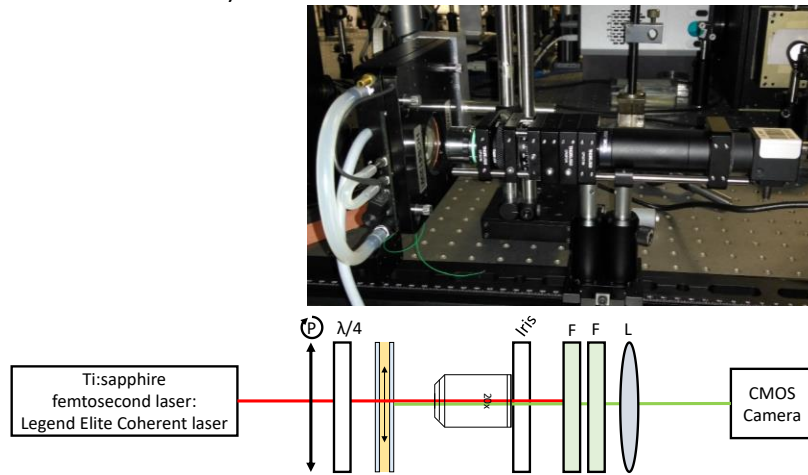

Figure SI.2. Picture and schematic representation of the SHG setup.

**B.3 Broadband Dielectric Spectroscopy**

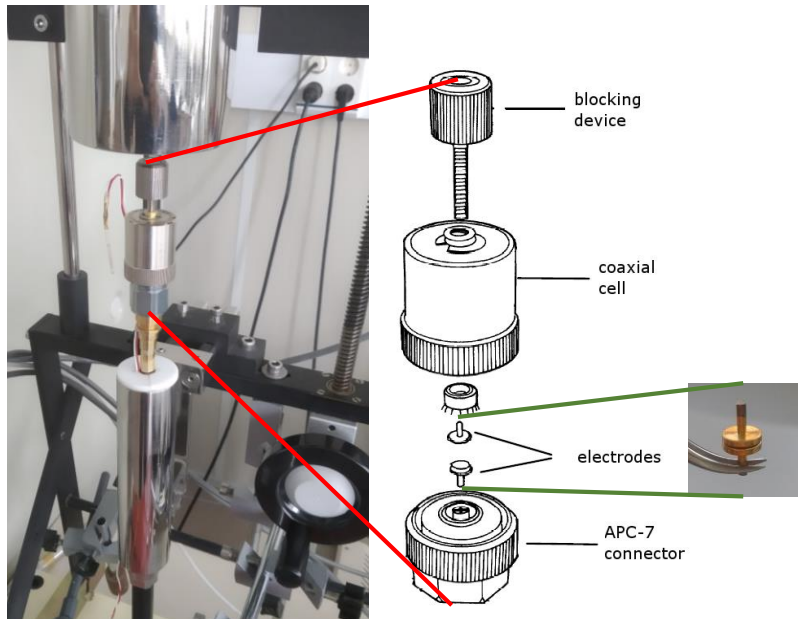

Figure SI.3. Picture and schematic representation of the sample holder (HP16091A) and sample capacitor for dielectric measurements.

Measurements of the complex permittivity $\varepsilon(\omega) = \varepsilon'(\omega) - i\varepsilon''(\omega)$ have been performed in the frequency range 10 Hz – 110 MHz combining two impedance analyzers (Alpha A from Novocontrol and HP4294). The measurement of the complex impedance at high frequencies requires of specialized setup. For this reason, the available sample holders have been modified in order to use the same sample cell with the different equipment needed to cover the frequency range. A modified HP16091A coaxial fixture, consisting of a sliding short circuit that can be moved along a 7 mm coaxial line section is used as sample holder. Within this setup small components can be placed between the sliding short and the central conductor. The sample is thus, a parallel plate capacitor made of two circular gold-plated brass electrodes of 5mm diameter and separated by 50 μm thick silica spacers (Fig. SI.3). Careful calibration of the sample plane with the different analyzers allows for nice overlap of the different frequency range measurements. The system is held in a Novocontrol cryostat and the spectra were recorded on cooling with different temperature steps being stabilized to ±50 mK. Measurements presented here were performed with the electrode surfaces untreated. From the comparison with results obtained for the electrodes treated for homeotropic alignment, those obtained in classical glass ITO cells for parallel alignment, and observations of the orientation of the *N* phase when in direct contact with the ITO surface, we can infer that in the bare gold electrodes the sample spontaneously aligns homeotropically in the nematic phase. Measurements were sequentially repeated for the different analyzers and two different cells, showing reproducibility within the experimental errors.

### C. Defect lines in the $N_s$ phase

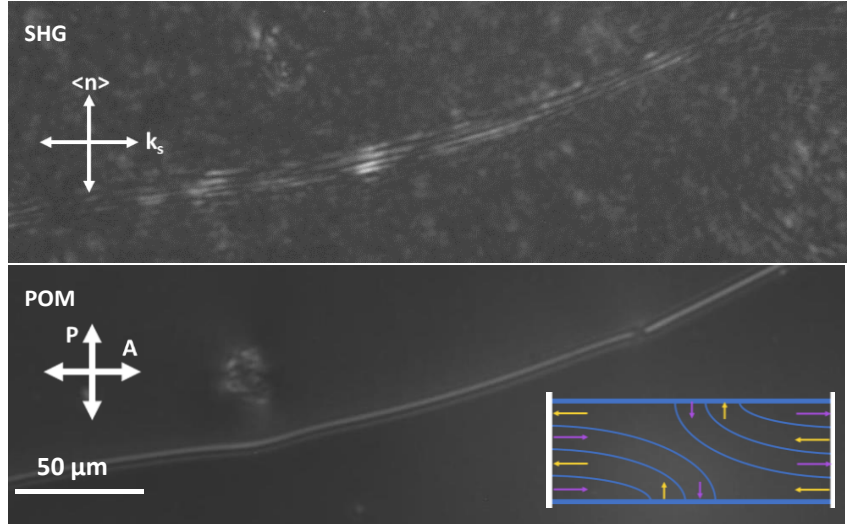

Fig. SI.4. Comparison of SHG and POM images around a defect line/wall at 123 °C. Several lines are observed in the SHG image, which suggest that the splay direction in the lines goes from out of plane to in plane as the schematic shown in the image. The grainy background signal in the SHG image can be partly attributed to coherent illumination (speckle) which results in inhomogeneous SHG signal generation and to inhomogeneities in the surfaces. In the POM image, white arrows indicate the direction of the polarizer and analyzer in each of the setups.

### D. Dielectric spectra and fits

Results of the temperature and frequency dependence of the real and imaginary components of the dielectric permittivity are shown in Fig. SI.4. in log-log scale. In the isotropic phase a single relaxation process is detected. In the nematic phase, the spectra is dominated by a relaxation process, whose frequency deviates from Arrhenius behavior close to the $N$-$N_s$ transition and whose strength strongly increases.

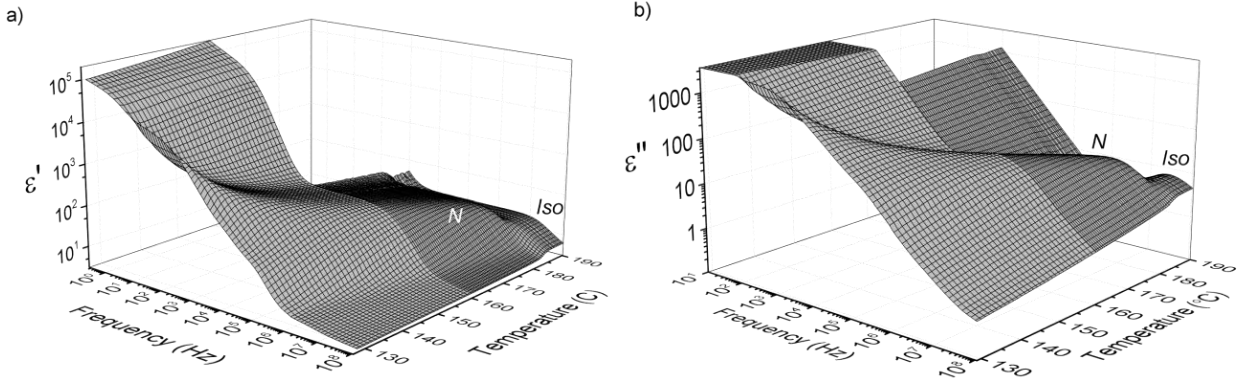

Fig. SI.5. 3D plots of the temperature and frequency dependence of the a) real $\varepsilon'$ and b) imaginary $\varepsilon''$ components of the dielectric permittivity.

For each temperature, the results were fitted to

$$\varepsilon(\omega) = \sum_k \frac{\Delta\varepsilon_k}{\left[1+(i\omega\tau_{HN})^{\alpha_k}\right]^{\beta_k}} + \varepsilon_\infty - i\frac{\sigma_0}{\omega\varepsilon_0} \qquad \text{SI.(3)}$$

SI

Where $\varepsilon_\infty$ is the high frequency permittivity, $\sigma_0$ is the dc-conductivity and each mode is fitted by the Havriliak-Negami (HN) function where $\Delta\varepsilon_k$ is the strength of each mode. The parameters $\alpha_k$ and $\beta_k$ ($0 < \alpha_k \leq 1$ and $0 < \beta_k \leq 1$) describe the broadness and symmetry of the relaxation spectra, respectively. The frequency of maximal loss is related to the shape parameters by the following equation [4]:

$$f_{max,k} = \frac{1}{2\pi\tau_{HN,k}} \left[\sin\left(\frac{\pi\alpha_k}{2+2\beta_k}\right)\right]^{1/\alpha_k} \left[\sin\left(\frac{\pi\beta_k\alpha_k}{2+2\beta_k}\right)\right]^{-1/\alpha_k} \qquad \text{SI.(4)}$$

As example of the procedure, Fig SI.5. shows the obtained fits for RM734 at one temperature of the isotropic phase, at one high temperature in the nematic range where only one process can be distinguished and one temperature close to the $N$-$N_s$ transition where 3 modes can be detected.

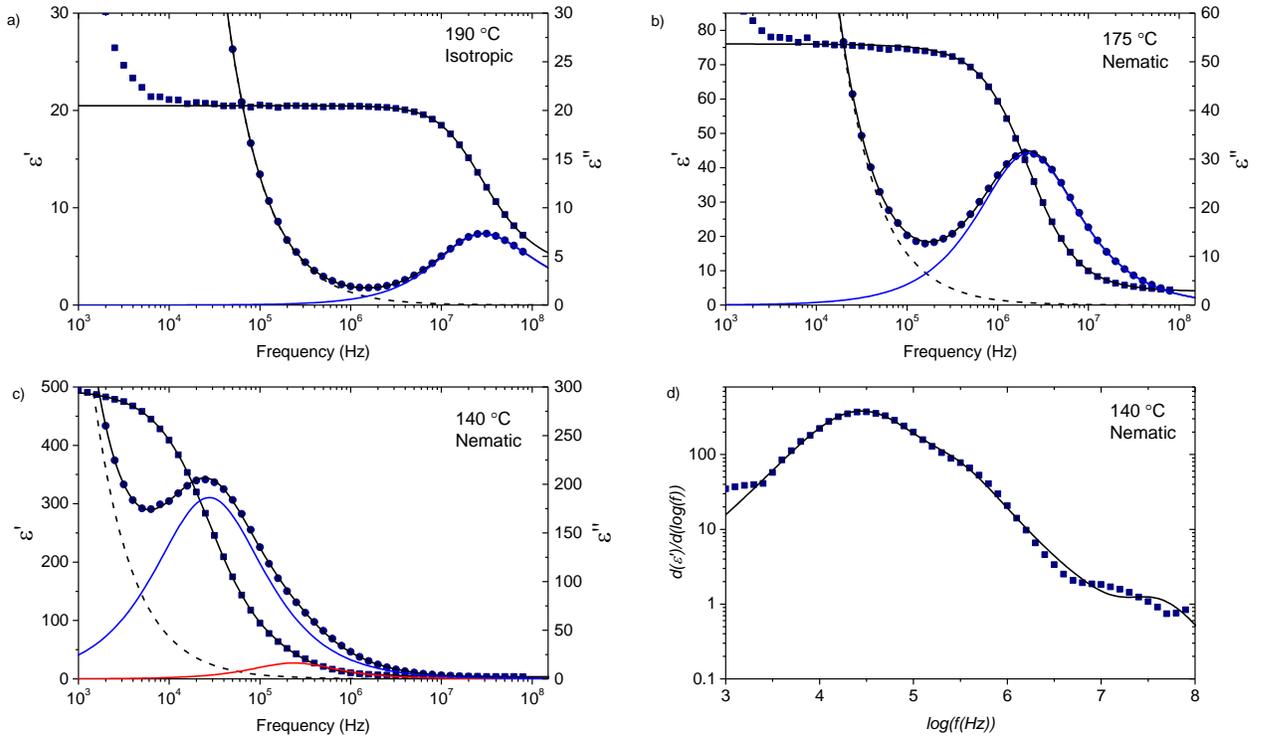

Fig. SI.6. (Color online) (a-c) Frequency dependence of the real (squares) and imaginary (circles) dielectric permittivity in the a) Isotropic phase and in the nematic phase at b) 175°C and c) 140°C. Solid lines result from fitting to Equation SI.1 and the corresponding deconvolution into the elementary processes. Dashed lines correspond to the current conductivity term. d) Derivative of the real part of the permittivity $d\varepsilon'/d(\log f)$ at 140°C for visualization of the relaxation modes.

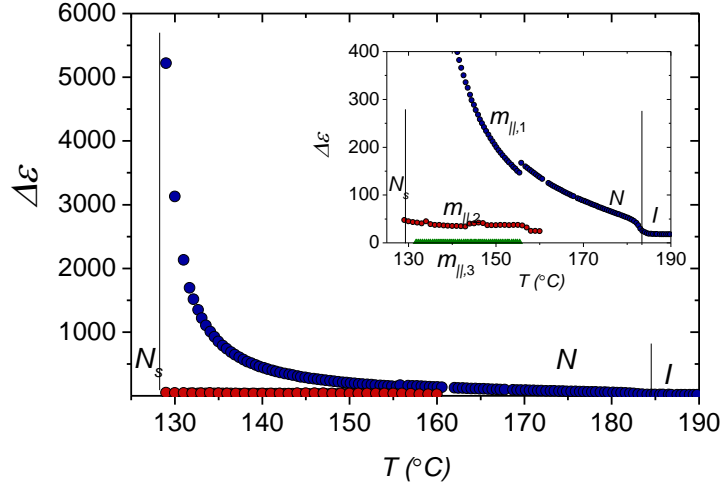

Fig SI.7. (Color online) Dielectric strength of the relaxation modes vs. temperature ($m_{\parallel,1}$ blue symbols, $m_{\parallel,2}$ red symbols and $m_{\parallel,3}$ green symbols)

### E. Coupled Splay and polarization fluctuations in the nematic phase

Orientational fluctuations are fundamental thermal excitations of the director field with two branches of eigenmodes: splay-bend and twist-bend [5]. In the case the director is coupled to another order parameter, the number of fluctuation modes increases. In the following, we will focus on pure splay fluctuations with wave vectors perpendicular to the director **n**. The fluctuation modes can be calculated from the free energy density,

$$f = \tfrac{1}{2} K_1 (\nabla \cdot \mathbf{n})^2 + \tfrac{1}{2} K_3 (\mathbf{n} \times (\nabla \times \mathbf{n}))^2 - \gamma \mathbf{n}(\nabla \cdot \mathbf{n}) \cdot \mathbf{P} + \tfrac{1}{2} t \mathbf{P} \cdot \mathbf{P} + \tfrac{1}{2} b (\nabla \mathbf{P})^2.$$  SI.(5)

Here, the first two terms are the usual Frank elastic free energy terms, where $K_1$ and $K_3$ are the splay and the bend orientational elastic constants, respectively. The third term describes the coupling between the splay deformation and electric polarization, where $\gamma$ is a bare splay flexoelectric coefficient. The last two terms are the lowest by symmetry allowed terms in **P** and $\partial P_i / \partial x_j$. In our case, the molecular dipole moment is along the molecular long axis, so $\mathbf{P} = P \mathbf{n}$.

We assume that the director is homogeneously aligned along the x-axis and, consequently, the average <P> is zero. The fluctuations can be described by a small angle $\varphi$ between the fluctuating director $\mathbf{n}(z,\tau) = (\cos(\varphi(z,\tau)), 0, \sin(\varphi(z,\tau)))$ and the average director, and normalized $P_n = P/P_r$, where $P_r$ is an arbitrary normalization constant. Here $\tau$ denotes time. The renormalized coefficients are: $t_r = t P_r^2$, $\gamma_r = \gamma P_r$, and $b_r = b P_r^2$. The free energy density (Eq. SI.3) expanded to the second order in $\varphi$ and $P$ simplifies to

$$f_{fl} = \frac{1}{2}\left( t_r P_n^2 + b_r \left(\frac{\partial P_n}{\partial z}\right)^2 - 2\gamma_r P_n \frac{\partial \varphi}{\partial z} + K_1 \left(\frac{\partial \varphi}{\partial z}\right)^2 \right),$$  SI.(6)

From which the linear coupled dynamic equations for $\varphi$ and $P_n$ follow

$$\eta_1 \frac{\partial \varphi}{\partial \tau} = K_1 \frac{\partial^2 \varphi}{\partial z^2} - \gamma_r \frac{\partial P_n}{\partial z},$$  SI.(7)

$$\eta_{P0}\frac{\partial P_n}{\partial \tau} = b_r \frac{\partial^2 P_n}{\partial z^2} - \gamma_r \frac{\partial \varphi}{\partial z} + t_r P_n \qquad \text{SI.(8)}$$

Here $\eta_1$ is the effective splay orientational viscosity and $\eta_{P0}$ the dissipation coefficients for $P$. The two solutions of Eqs. (SI.5) and (SI.6) are: $\varphi_{1,2} = \pm\varphi_{01,02}\sin(qz)e^{-\tau/\tau_{01,02}}$ and $P = P_{01,02}\cos(qz)e^{-\tau/\tau_{01,02}}$, with the relaxation rates

$$\frac{1}{\tau_{01,02}} = \frac{b_r\eta_1 q^2 + \eta_P K_1 q^2 + \eta_1 t_r \mp \sqrt{\left(b_r^2\eta_1^2 q^4 + 2b_r\eta_1 q^2\left(\eta_1 t_r - \eta_{P0}K_1 q^2\right) + \eta_{P0}^2 K_1^2 q^4 - 2\eta_1\eta_{P0}K_1 q^2 t_r + 4\gamma_r^2\eta_1\eta_{P0}q^2 + \eta_1^2 t_r^2\right)}}{2\eta_1\eta_{P0}}$$

SI.(9)

For small $q$ they simplify to

$$\frac{1}{\tau_{01}} = \frac{\left(K_1 - \frac{\gamma^2}{t}\right)}{\eta_1} q^2 \qquad \text{SI.(10)}$$

$$\frac{1}{\tau_{02}} = \frac{t}{\eta_P} + \left(\frac{b}{\eta_P} + \frac{\gamma^2}{\eta_1 t}\right)q^2 \qquad \text{SI.(11)}$$

Here, $\eta_P = \eta_{P0}P_r^2$.

The first mode is of a hydrodynamic type and is mainly a director mode. If the splay elastic constant is replaced by an effective elastic constant

$$K_{1,eff} = K_1 - \frac{\gamma^2}{t}, \qquad \text{SI.(12)}$$

the relaxation rate Eq.(6) becomes identical to the relaxation rate of the splay fluctuations [5] with the splay elastic constant replaced by the effective one. This shows that in a usual measurement of the splay elastic constant by either dynamic light scattering (DLS) or by Frederiks transition, always the effective elastic constant Eq. (SI.10) is measured. The second mode is of an optic type with relaxation rate which is finite at $q = 0$ and this is the collective mode observed in the dielectric spectroscopy.

The average square amplitudes of the fluctuation modes can be calculated by using equipartition theorem. For small $q$ the amplitudes of the director fluctuations are given as

$$\left\langle \varphi_{01}^2 \right\rangle = \frac{k_B T}{V K_{1,eff} q^2} \qquad \text{SI.(13)}$$

$$\left\langle \varphi_{02}^2 \right\rangle = \frac{2 k_B T \gamma^2 \eta_{P0}^2}{V t^3 \eta_1^2} q^2 \qquad \text{SI.(14)}$$

Here $V$ is the volume of the sample. The first mode is much slower and has much larger amplitude, and is the one measured in the DLS experiment.

The average square amplitudes of the polarization at $q = 0$ are

$$\left\langle P_{01}^2 \right\rangle = \frac{2 k_B T \gamma^2}{V K_{1,eff} t^2} \qquad \text{SI.(15)}$$

$$\langle P_{02}^2 \rangle = \frac{2k_B T}{Vt} \qquad \text{SI.(16)}$$

In the dielectric experiment, the modes at *q* = 0 and with finite relaxation rates are measured, which in our case is the second mode.

The amplitude of the collective mode measured in the dielectric spectroscopy is proportional to $\langle P_{02}^2 \rangle$ [6]

$$\Delta \varepsilon_{\Box 1} \propto \langle P_{02}^2 \rangle \propto 1/t . \qquad \text{SI.(17)}$$

In the Landau description of the ferroelectric phase transition the coefficient in the quadratic term of **P** is inverse of the electric susceptibility. Eqs. (SI.14) and (SI.15) shows that.in our case, the contribution of the dielectric susceptibility that drives the ferroelectric phase transition is the second mode and

$$t = \frac{1}{\varepsilon_0 \Delta \varepsilon_{\Box 1}} , \qquad \text{SI.(18)}$$

where $\varepsilon_0$ is the vacuum permittivity.